\documentclass[10pt,conference]{IEEEtran}
\IEEEoverridecommandlockouts
\usepackage{cite}
\usepackage{booktabs}
\usepackage{multirow} 
\usepackage{pifont}
\usepackage{graphicx} %
\usepackage{makecell}
\usepackage{adjustbox}
\usepackage{subcaption}
\usepackage{tablefootnote}
\usepackage{threeparttable}
\usepackage[most]{tcolorbox}
\usepackage{url}
\usepackage{footmisc}

\usepackage{colortbl}
\usepackage{color}
\usepackage{xcolor}
\usepackage{makecell}
\usepackage{enumitem}
\usepackage{graphicx}
\usepackage{xcolor}
\usepackage{amsmath,amssymb,amsfonts}
\usepackage{algorithmic}
\usepackage{graphicx}
\usepackage{textcomp}

\usepackage{booktabs}
\usepackage{tabularx}
\usepackage{array}
\usepackage{multirow}
\def\BibTeX{{\rm B\kern-.05em{\sc i\kern-.025em b}\kern-.08em
    T\kern-.1667em\lower.7ex\hbox{E}\kern-.125emX}}

\usepackage{microtype}
\definecolor{lightTeal}{HTML}{add2ca}
\definecolor{softPurple}{HTML}{8c89ba}
\definecolor{paleGreen}{HTML}{e6f0db}
\definecolor{mutedOrange}{HTML}{dfa660}
\definecolor{limeGreen}{HTML}{97c04b}
\definecolor{lightBlue}{HTML}{dbeef3}
\definecolor{softRed}{HTML}{d77470}
\definecolor{dustyBlue}{HTML}{93b2d9}
\definecolor{tabletitle}{HTML}{E5DBF8} 

\definecolor{checkgreen}{HTML}{009900}
\definecolor{crossred}{HTML}{CC0000}

\AtBeginDocument{%
  \providecommand\BibTeX{{%
    Bib\TeX}}}

\usepackage{mdframed}
\newcommand{\answerRQ}[2]{
    \vspace{1mm}
    \begin{mdframed}[
        linecolor=gray,
        roundcorner=12pt,
        backgroundcolor=gray!15,
        linewidth=3pt,
        innerleftmargin=2pt,
        leftmargin=0cm,
        rightmargin=0cm,
        topline=false,
        bottomline=false,
        rightline=false
    ]
    \textbf{Findings #1:} #2
    \end{mdframed}
}

\usepackage{hyperref}
\hypersetup{colorlinks = true,
	linkcolor = black,
	urlcolor = black,
	citecolor = black}

\begin{document}

\title{How Agent Skills Evolve: A Large-Scale Empirical Study of Skills on GitHub}

\author{

\IEEEauthorblockN{Jinjun Huang\IEEEauthorrefmark{2}, Jialun Cao\IEEEauthorrefmark{3}, Junxiao Han\IEEEauthorrefmark{4}, Zhonghao Jiang\IEEEauthorrefmark{2}, Shing-Chi Cheung\IEEEauthorrefmark{3}, Zhongxin Liu\IEEEauthorrefmark{2}\IEEEauthorrefmark{1}\thanks{\IEEEauthorrefmark{1}Corresponding author.}}
\IEEEauthorblockA{\IEEEauthorrefmark{2}Zhejiang University, Hangzhou, China\\
\IEEEauthorrefmark{3}The Hong Kong University of Science and Technology, Hong Kong, China\\
\IEEEauthorrefmark{4}Hangzhou City University, Hangzhou, China\\
Emails: \{huanghuanghuang, zhonghao.j, liu\_zx\}@zju.edu.cn, \{jialuncao, scc\}@cse.ust.hk, hanjx@hzcu.edu.cn}
}

\maketitle

\begin{abstract}

Large language model-based agents increasingly rely on external skills to reuse task-specific knowledge without retraining the underlying model. 
As skills become a common practice for extending agent capabilities, understanding creation and evolution is critical for building reliable agent systems.
However, existing studies mainly examine agent skills from a static perspective, leaving their evolution in real-world repositories largely unexplored.
This limits our understanding of the underlying patterns and latent motivations behind changes in skills.
To fill this gap, this paper presents a large-scale empirical study of agent skills, capturing 84,200 skills and 750,898 related commits from GitHub.
We manually construct taxonomies of skill components, change patterns, and evolution motivations, and build an automated annotation pipeline to label skill changes at scale.
Based on the taxonomy, we conduct quantitative analyses of current skill composition and post-creation evolution frequency, together with component-level qualitative assessments of modified components, change patterns, and motivations.
Our findings reveal five key implications:
(i) reliability receives less attention in skill composition and change, motivating execution-contract skill design that incorporates verification and recovery mechanisms; 
(ii) evolution changes concentrate in project-coupled skills, calling for automated just-in-time skill updating mechanisms; 
(iii) change patterns are dominated by incremental enrichment and refinement, motivating pattern-guided skill enhancers that diagnose and improve incomplete skill components;
(iv) redundancy is a practical maintenance issue, highlighting the need for context-efficient skill representation;
(v) security-motivated changes show limited risk reduction, suggesting skill regression testing methods that objectively verify skill functionality and security.
Beyond these insights, we envision our dataset to provide a basis for future studies and work on automated skill-related techniques.

\end{abstract}

\section{Introduction}

\begin{figure}[!t]
  \centering
  \includegraphics[width=1.0\linewidth]{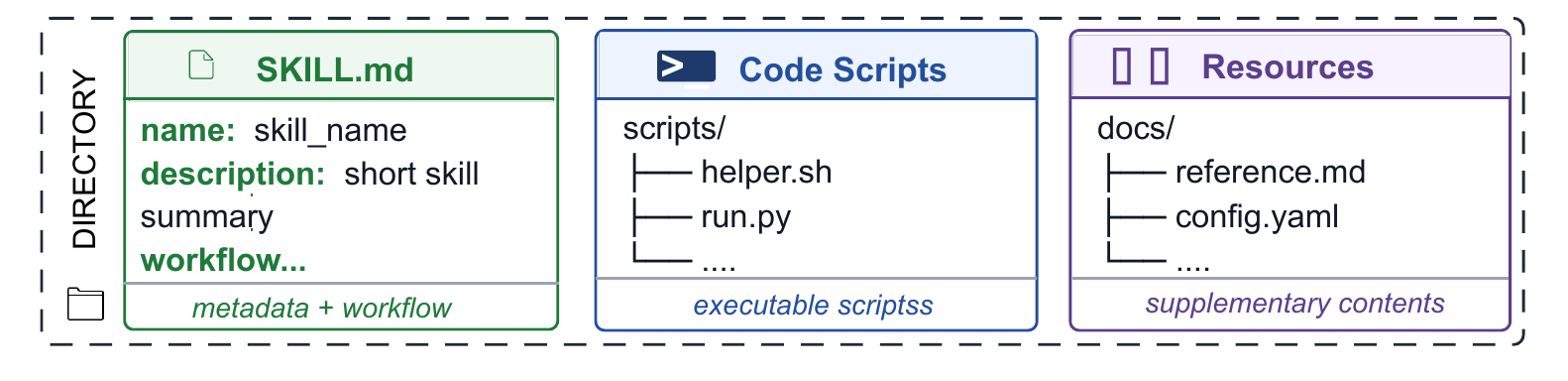}
  \caption{File Structure of an Agent Skill} 
  \label{fig:1_skill_case}
\end{figure}

Recent artificial intelligence (AI) systems have been shifting from single-turn, passive, reactive large language models (LLMs) to autonomous AI agents~\cite{11103638, yao2022react,xu2026theagentcompany,liu2024agentbench}.
Unlike traditional LLM systems that primarily generate responses to user inputs~\cite{10172803,10172800,kang2023large}, AI agents are designed to accomplish task objectives autonomously: they can perform multi-step reasoning and planning, invoke external tools on demand, interact with working environments, and maintain task context through memory mechanisms~\cite{bouzenia2025understanding,du2025memr,xu2025mem,tan2025prospect}.
Among these capabilities, \textit{agent skills} are introduced as an important mechanism for packaging reusable task knowledge and execution procedures.
A skill is a structured, reusable functional unit that can be loaded into the agent context as needed~\cite{anthropic2025agentskills}.
As illustrated in Figure~\ref{fig:1_skill_case}, a skill is typically organized around a main instruction file (e.g., \texttt{SKILL.md}), which contains metadata for retrieval, workflow instructions, and task-specific knowledge, and may be accompanied by executable scripts and associated resources.
By retrieving and composing skills during execution, agents can acquire domain-specific capabilities without retraining the underlying model, making skills an important building block for extending agent functionality~\cite{su2026skill, yang2026autoskill}.

With models, tools, memory, and external knowledge (skills) orchestrated around agents, agent-based systems are evolving from prompt-response interfaces into a new form of software systems.
In this form, foundation models serve as decision-making cores, tools provide external actions, memories preserve task context and experience, and skills encapsulate reusable execution knowledge.
From a software engineering perspective, agent skills can be viewed as a new class of software artifacts.
Skills are created, revised, reused, and maintained. %
They need to be continuously updated to changing task requirements, tool interfaces, execution environments, and user feedback.
Studying these updates helps understand how skills evolve in practice and provides empirical guidance for writing, validating, and improving skills in the future.

The rapid evolution of the ecosystem surrounding agent skills has attracted increasing research attention\cite{xia2025demystifying,bouzenia2025you}.
Empirical studies have been conducted to examine the acquisition, retrieval, and security risks of skills~\cite{xu2026agent,ling2026agent,liu2026agent}.
While these studies provide a static view of skills, the longitudinal evolution of real-world skill artifacts is under-explored, i.e., how they change across time is still unknown.
This gap limits our understanding of the underlying patterns and latent motivations behind changes in skills.
To bridge the gap, we construct a large-scale dataset of 84,200 agent skills and 750,898 related commits from real-world GitHub repositories and conduct an empirical study of skill evolution.
By studying skills as evolving software artifacts, we aim to learn how frequently they evolve, which parts are modified, what kinds of changes developers make, and what motivations drive these changes.

We study the following research questions and sub-questions.
    
\begin{itemize}[left=0pt, topsep=0em]

\item \textbf{RQ1: What is the current status of agent skills in open-source repositories?}  
We provide an overview of existing skills, focusing on their composition and evolution frequency.

\begin{itemize}[leftmargin=1.5em]
    \item \textbf{RQ1.1: What are the main components of agent skills?}  
    We manually construct a component taxonomy from inspected samples.
    Based on this taxonomy, we classify skill content into fine-grained functional components and quantify their occurrence in real-world repositories.
    The results show that most skills are currently used primarily to structure prompts, procedural guidance, and domain knowledge.
    At the fine-grained level, we observe that skills provide limited explicit quality assurance and failure-handling mechanisms.
    
    \item \textbf{RQ1.2: How frequently do agent skills evolve over time?}  
    We identify content-related modification commits as evolution events and analyze their counts and temporal intervals.
    Our findings show that most skills experience limited maintenance, and frequent evolution is concentrated in a small subset of complex and project-coupled skills.
    This concentration suggests that project-coupled skills incur high manual maintenance costs as their dependent projects evolve, highlighting the need for automated mechanisms that can efficiently propagate project-level changes into corresponding skill updates.
    
\end{itemize}

\item \textbf{RQ2: Which components of agent skills are modified during evolution?}  
Based on the collected dataset, we retrieve historical skill versions, parse file-level diffs, and use LLM-assisted component labeling to map changed content to functional components.
We find that skill evolution is concentrated in instruction-oriented components. 
Within the instruction-oriented components, reliability-oriented components often receive less attention from developers.
Given the inherent ambiguity of skills as natural language artifacts and the stochastic nature of LLM-based execution, the outcomes can be highly unpredictable. Reliability is therefore a non-negligible concern for developers.

\item \textbf{RQ3: What types of changes characterize agent skill evolution?}  
To organize skill changes and their motivations under a unified framework, we sample evolution commits and manually construct a taxonomy of change patterns and latent motivations.
Guided by the taxonomy, we aggregate file-level diffs into changed-component units, use LLM-assisted labeling to assign each unit a \textit{how} label, and quantify the resulting distribution of change patterns.
The results indicate that change patterns is dominated by incremental enrichment and refinement, while structural reorganization is rare.
This suggests that future automated skill maintenance tools may prioritize iterative enrichment- and refinement-oriented support over structural-level revisions.
Redundancy is also a practical maintenance issue in skill evolution, with developers frequently simplifying skill content by removing unnecessary artifacts and duplicated instructions.
This is especially beneficial for token saving in long-horizon agent task execution, where repeated skill invocations can otherwise lead to substantial context accumulation.

\item \textbf{RQ4: What motivates the evolution of agent skills?}  
Using the motivation dimension of the taxonomy constructed in RQ3, we also annotate each changed-component unit with a label and analyze the distribution of motivations.
This analysis shows that skill evolution is dominated by Perfective maintenance, especially improving functionality and retrievability.
In contrast, reliability- and risk-oriented maintenance receive limited attention.
As AI agents frequently involve substantial risky operations, such as file modification and network access, ensuring the reliability and risk-prevention of skills that govern these operations should be treated as a maintenance priority.
For security-related changes, we further compare the scanner-reported risks before and after the corresponding commits, and find that only a small fraction of changes are accompanied by observable reductions in risks.
This gap between intended and observed risk reduction highlights the need for executable tests that can objectively verify skill functionality and security.

\end{itemize}

To the best of our knowledge, we are the first large-scale empirical study of agent skill evolution, providing a comprehensive understanding of the current status, evolution patterns, and motivations behind skill changes. In summary, this paper makes the following contributions: 

\begin{itemize}[left=0pt, topsep=0em]
\item \textbf{Dataset.} We construct and release a large-scale dataset capturing the evolution history of agent skills, comprising 84,200 skills and 750,898 commits from real-world GitHub repositories. This dataset provides a foundation for a wide range of empirical analysis.
\item \textbf{In-depth Study.} We conduct a comprehensive empirical study of agent skill evolution, uncovering characteristics, key patterns, and motivations behind skill changes.
\item \textbf{Implications.} Based on our findings in skill evolution, we
discuss research opportunities in the skill ecosystem, focusing on automated skill refinement, skill testing, and context-efficient skill representation.

\end{itemize}

\section{Background and Related Works}

\subsection{Agent skills: modular and reusable capabilities}
\label{sec:2_skill_background}

An agent skill is a predefined, reusable, task-oriented capability module that helps agents solve specific tasks.
Unlike knowledge encoded in model parameters, skills are external artifacts that explicitly describe task-specific workflows, tool usage patterns, constraints, and domain expertise.
This design allows agents to acquire new capabilities without retraining, reuse prior task experience across similar scenarios, and improve the consistency and reliability of task execution\cite{zhao2024expel,zhang2024offline,wang2025agent}.
Structurally, a skill is typically organized as a standalone directory.
As shown in Figure~\ref{fig:1_skill_case}, it usually contains a core guidance file, such as \texttt{SKILL.md}, which includes retrieval metadata and execution workflow, and may include executable scripts or associated resources such as reference documents and configuration files.

Recent studies divide the lifecycle of agent skills into three stages: acquisition, retrieval/selection, and evolution~\cite{zhou2026comprehensive}. A typical lifecycle starts with skill acquisition, where reusable procedural artifacts are created from human expertise or execution traces. 
These skills are then indexed and retrieved from skill libraries, selected according to the current task context. 
Based on execution feedback, existing skills can be modified and evolved over time.
Following this lifecycle view, existing techniques can be grouped into skill acquisition, skill retrieval and selection, and skill evolution.
Prior work addresses skill acquisition from interaction experience or external documents~\cite{liang2026skillnet,li2026organizing,ni2026trace2skill}, skill retrieval and selection based on task context and constraints~\cite{zheng2026skillrouter,liu2026graph,zeng2026group}, and skill evolution through failure-driven revision~\cite{pu2026skillops,zhang2026coevoskills,li2026skillhone}.
Rather than proposing a specific technique, our work empirically studies real-world skill evolution from multiple perspectives.

\subsection{Empirical Studies on Agent Skills}

As agent skills have rapidly emerged in agent ecosystems, recent studies have begun to empirically characterize their content, structure, and security~\cite{liu2026agent, liu2026malicious, ling2026agent}. 
Ling et al.~\cite{ling2026agent} conduct a large-scale data-driven analysis of publicly listed Claude skills, examining their categories, adoption patterns, length distribution, and redundancy.
Liu et al.~\cite{liu2026agent} further study the security risks of agent skills at scale.
They identify vulnerability patterns such as prompt injection, data exfiltration, privilege escalation, and supply-chain risks. 
Complementary to vulnerability analysis, another recent study~\cite{liu2026malicious} constructs a behaviorally verified dataset of malicious skills and reveals how malicious instructions, hidden behaviors, and attack payloads appear in real-world skill ecosystems.

However, they mainly treat skills as static artifacts and focus on characteristics of content, while the longitudinal evolution of real-world skills is under-explored. 
Our work complements them by reconstructing skill evolution histories from commits and analyzing modified components, evolution patterns, and underlying motivations behind these changes.

\section{Methodology}

\begin{figure*}[htbp]
  \centering
  \includegraphics[width=0.85\textwidth]{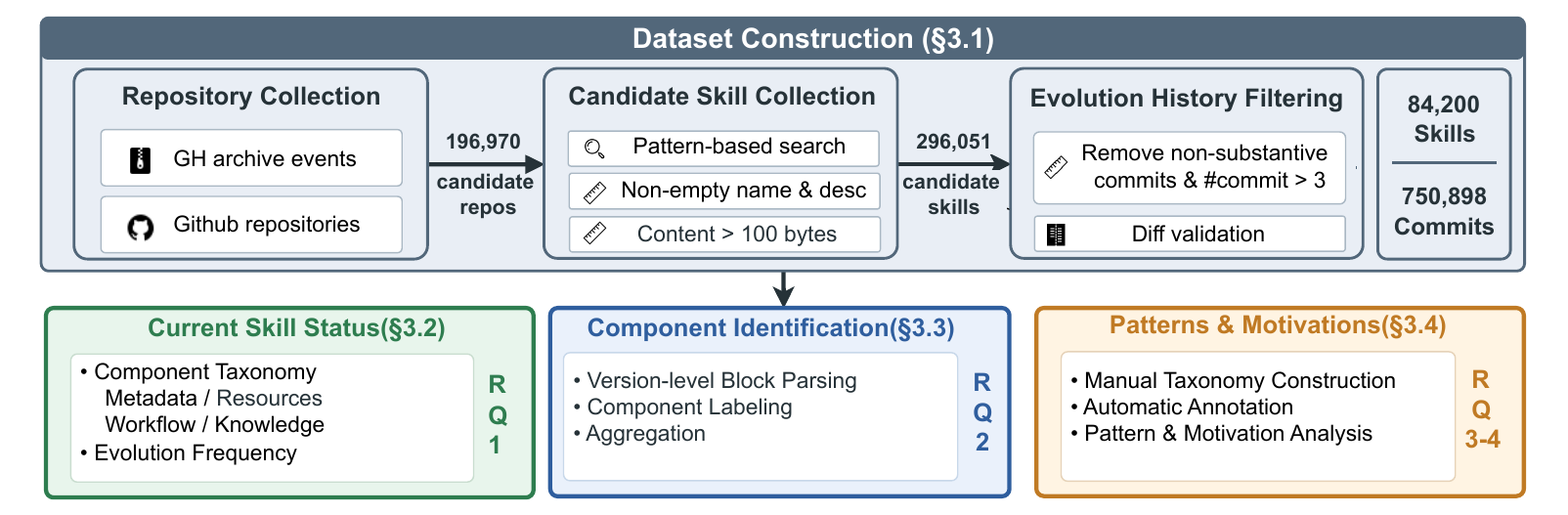}
  \caption{Overview of Methodology} 
  \label{fig:3_overall_method}
\end{figure*}

In this section, we present the methodology of our study. Our pipeline is described in Figure~\ref{fig:3_overall_method}, from dataset collection to the analysis of collected data.

\subsection{Dataset Construction}
\label{sec:3_dataset_construction}
To investigate how agent skills evolve in real-world settings, a dataset that contains skills with evolutionary histories is needed.
This section describes our dataset construction pipeline, including repository collection, candidate skill collection, and evolution history filtering.

\noindent\textbf{Repository Collection.}
GitHub hosts over 300 million repositories. 
The first challenge lies in identifying high-quality repositories that may contain skill files.
As the concept of agent skills was formally introduced by Anthropic in October 2025~\cite{anthropic2025agentskills}, we conservatively set September 1, 2025, as the starting date of our collection window, which extends through April 23, 2026.
We leverage the GH Archive dataset~\cite{gharchive} to identify active and potentially relevant repositories.
Specifically, we retain repositories with more than three stars and more than three push events during the target period. 
We use these low thresholds as a conservative lower bound, ensure a minimal level of public visibility and recent development activity while retaining newly created repositories that may contain agent skills.
This stage provides the initial pool of 196,970 repositories for subsequent skill file mining.

\noindent\textbf{Candidate Skill Collection.}
Based on the collected repositories, we identify files that potentially correspond to agent skills.
Since skill files may follow different naming conventions and directory layouts across projects, we adopt a pattern-based search strategy.
Specifically, we consider files matching common skill-related patterns, including \texttt{skill.md}, files under \texttt{.codex/skills} and \texttt{.claude/skills}, and Markdown files located in directories named \texttt{skills}.
Following existing skill file specifications~\cite{agentskills2026agentskills}, we further apply heuristic rules to remove clearly invalid candidates.
Specifically, we retain a skill file only when it provides the required metadata, including non-empty \texttt{name} and \texttt{description} fields, and contains sufficient content, i.e., more than 100 bytes.
This stage yields 296,051 candidate skill files.

\noindent\textbf{Evolution History Filtering.}
Finally, we filter the evolution histories of the collected candidate skills. 
To ensure that each skill has an analyzable evolutionary history, we require each retained skill file to have more than three commits.
Here we define non-substantive commits as commits that are explicitly intended to fix minor textual errors.
Concretely, we identify potentially non-substantive commits based on commit messages containing keywords such as \texttt{typo}, \texttt{spelling}, and \texttt{grammar}. 
Then we validate whether the remaining commits contain content changes by checking their file-level patches, retaining only commits with non-whitespace added or removed lines.
After deduplication, we obtain a dataset composed of 84,200 skills across 7,776 repositories, along with 750,898 related commits.

\subsection{Analysis of Current Skill Status}

To address RQ1, we conduct a statistical analysis based on the dataset from two perspectives: structural composition and evolution frequency.

\noindent\textbf{Detailed Component Category.} 
\label{sec:component_label_method}
As discussed in \ref{sec:2_skill_background}, the existing open format for agent skills mainly categorizes skill content by file type~\cite{agentskills2026agentskills}. 
However, file-level categorization cannot capture the evolution of fine-grained functional contents within a skill. 
Therefore, we further classify skill content from a functional perspective to support fine-grained evolution analysis.
 
To determine these categories, 383 skill files are independently and manually labeled by two authors of the study. 
Following prior work\cite{10.1145/3715777,charoenwet2024empirical}, the sample size ensures a 95\% confidence level with a $\pm$5\% margin of error.
After the independent data annotation process, a collaborative review was undertaken by the two authors to reconcile disagreements and finalize the categories through consensus, thereby solidifying the taxonomy used in our subsequent analysis.
Finally, we established five top-level categories: \textit{metadata}, \textit{workflow}, \textit{knowledge}, \textit{resources}, and \textit{others}, which are shown in Table~\ref{tab:3_skill_component_taxonomy}.

\begin{table*}[t]
\centering
\caption{Taxonomy of skill components.}
\label{tab:3_skill_component_taxonomy}
\renewcommand{\arraystretch}{1.15}
\setlength{\tabcolsep}{5pt}
\begin{tabularx}{\textwidth}{
    >{\raggedright\arraybackslash}p{0.18\textwidth}
    >{\raggedright\arraybackslash}p{0.20\textwidth}
    >{\raggedright\arraybackslash}X
}
\toprule
\textbf{Component} & \textbf{Subcomponent} & \textbf{Definition} \\
\midrule

\textbf{Metadata}
& Identity
& Basic information that identifies a skill, such as name, version, author, or category. \\

& Functionality
& Descriptions of the skill's capability. \\

& Trigger condition
& Conditions that determine when the skill should be retrieved or invoked. \\

\midrule

\textbf{Workflow}
& Execution
& Step-by-step procedures and operational logic to obtain results \\

& Verification
& Steps for validating intermediate or final results during the workflow. \\

& Recovery
& Error-handling or fallback strategies for failed executions. \\

& Format specification
& Specifications of inputs, outputs, and intermediate results \\

\midrule

\textbf{Knowledge}
& Internal guidance
& Background knowledge or domain-specific guidance embedded in \texttt{Skill.md}. \\

& Complementary documents
& External reference documents that supplement the main skill file. \\

\midrule

\textbf{Resources}
& Code scripts
& Executable scripts or program files used during task execution. \\

& Data files
& Auxiliary data files, such as configurations, credentials, or dependencies. \\

\midrule

\textbf{Others}

& Miscellaneous
& Other files that do not fall into the above categories, like evaluation files. \\

\bottomrule
\end{tabularx}
\end{table*}

\noindent\textbf{Evolution Frequency Overview.}
To characterize the overall activeness of skill evolution, we reconstruct each skill's post-creation evolution history from repository commits.
We define a commit that modifies a specific skill as a \textit{skill evolution event}, and exclude each skill's initial creation commit, as our focus is on post-creation evolution rather than skill introduction.
Based on these events, we measure: (1) the number and proportion of skills with at least one evolution event, (2) the number of evolution events per skill, and (3) the time intervals between consecutive evolution events.
These metrics capture the prevalence, frequency, and temporal rhythm of skill evolution.

\subsection{Identification of Modified Skill Components}
\label{sec:3_component_identification}

To address RQ2, we further investigate where changes occur within skill content.
Specifically, we take each \textit{skill evolution event} as the unit of analysis and map the modified contents in each event to the predefined skill components based on the previously constructed component taxonomy. 
To illustrate this process, we provide an example in our online appendix\footnote{\label{fn:online_appendix}\url{https://github.com/ZJU-CTAG/SkillEMP}}.

\begin{figure*}[t]
  \centering
  \includegraphics[width=1.0\linewidth]{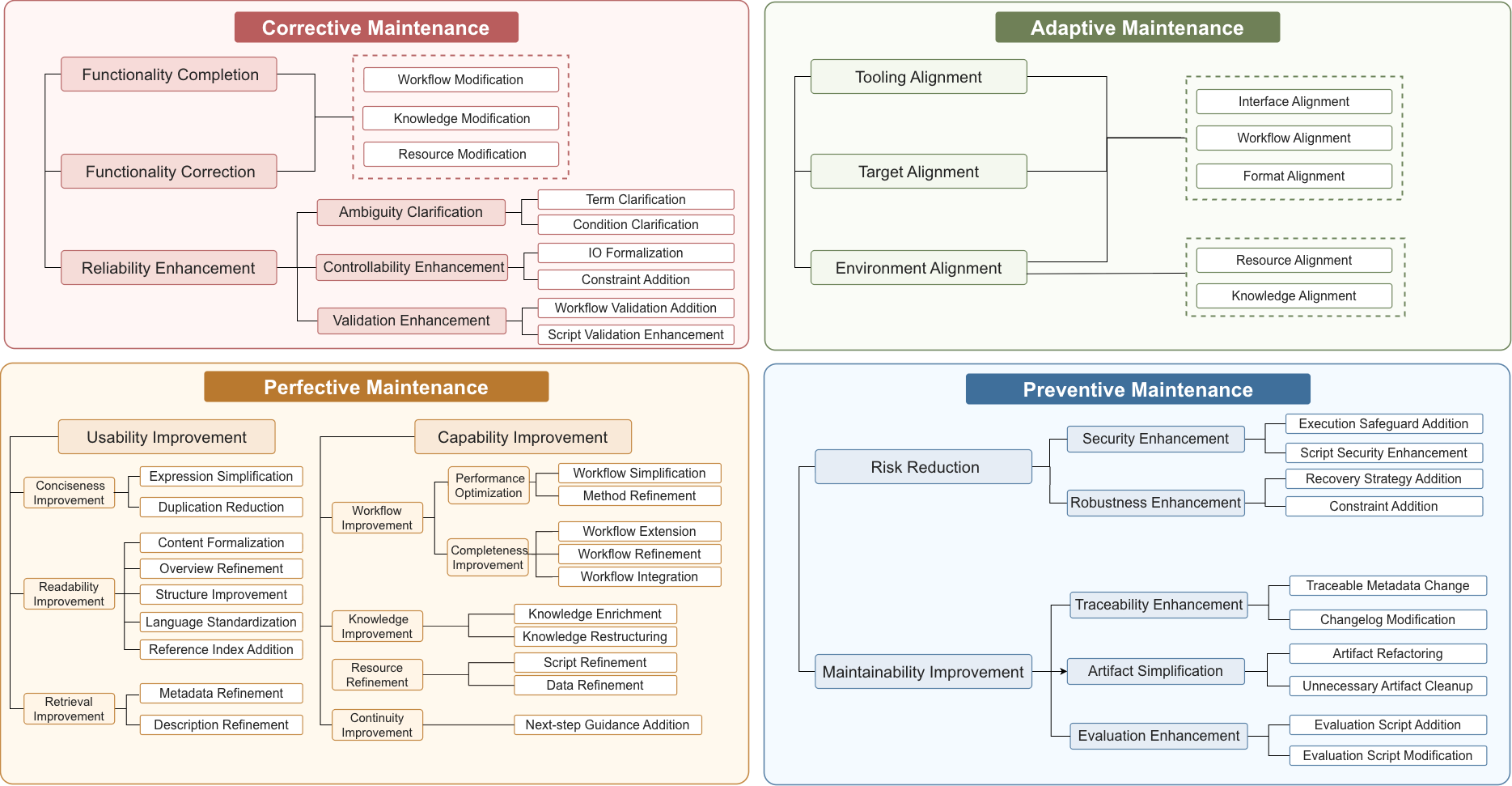}
  \caption{Taxonomy of skill evolution motivations and patterns. Colored nodes denote fine-grained ``why'' categories, while white nodes denote concrete ``how''.}
  \label{fig:3_taxonomy_evolution}
\end{figure*}

\noindent\textbf{Step 1: Version-level Block Construction.} A skill directory typically contains a main instruction file (i.e., \texttt{SKILL.md}) and several auxiliary resource files, such as scripts, data files, and supplementary documents. Since these two types of files play different roles in a skill, we process them separately.

For the main instruction file, we retrieve both its pre-change and post-change versions for each evolution event.
We parse each version into blocks based on Markdown format, including frontmatter, paragraphs, lists, tables, and code fences. Each block preserves its line range, block type, heading path, and textual content. 
This version-level block representation allows us to assign semantic components to the skill content in a more structured manner.
For auxiliary resource files, we do not parse their internal content because these files are heterogeneous in format and lack a unified structure.
Instead, we extract their file-level diffs from the commit patch as evidence for analyzing their concrete modifications.

\noindent\textbf{Step 2: Component Labeling and Change Localization.} After constructing Markdown blocks for both versions of \texttt{SKILL.md}, we map these blocks to our predefined skill component taxonomy. 
Specifically, we provide the full content as context and ask an LLM to assign each target block to exactly one skill component. 
The main instruction file is classified into components, including metadata, workflow, and knowledge.
Auxiliary resource files are also categorized by file type, including knowledge and resources.
For labeling, we use \texttt{DeepSeek-V4-Flash}~\cite{xu2026deepseek}, as it provides strong semantic understanding capability at a reasonable cost for large-scale labeling. 
We evaluate the reliability of automated annotation in Section~\ref{dis:accuracy}, where the component-labeling task achieves an accuracy score of 0.881 against manually labeled data.

We then perform line-level analysis for the commit diff between the pre-change and post-change versions. 
For modifications, the removed side is mapped to blocks in the pre-change version, and the added side is mapped to blocks in the post-change version. 
Pure deletions and insertions are mapped to the pre-change and post-change versions, respectively. 
Each changed line is mapped to its corresponding labeled Markdown block, thereby identifying the affected skill component. 
For auxiliary resource files, we map the file-level change to the component label assigned to the corresponding file.
The detailed prompt for labeling is provided in our online appendix\footref{fn:online_appendix}.

\noindent\textbf{Step 3: Aggregation.} Finally, we aggregate the labeled changes at the granularity of a skill evolution event. 
If multiple changed Markdown blocks or auxiliary files in the same evolution event are mapped to the same component, we merge them into one changed component.
As a result, each evolution event is represented as a set of changed skill components. 
Based on this representation, we compute the frequency and proportion of modifications for each skill component.
This enables us to identify which parts of agent skills are more likely to evolve and provides the basis for further analyzing their evolution patterns and motivations.

\subsection{Taxonomy of Evolution Patterns and Motivations}
\label{sec:3_taxonomy_evolution_pattern_and_motivation}

To address RQ3 and RQ4, we construct a two-dimensional taxonomy of skill evolution on top of the previously annotated skill components. 
Specifically, the ``how'' dimension captures the change operation applied to a modified component, while the ``why'' dimension captures the evolution motivation behind the modification. 

\noindent\textbf{Taxonomy Construction.} 
We randomly sampled 384 commits to annotate, following the same independent annotation and consensus review procedure in ~\ref{sec:component_label_method}.
This sample size also satisfies the minimum requirement for a 95\% confidence level with a $\pm 5\%$ margin of error. 
To further assess the stability of the taxonomy, we additionally inspected 20 randomly selected commits after the initial taxonomy was constructed.
In total, our manual inspection covered 404 commits across 27 skills.
The additional inspection did not lead to any changes to the taxonomy, indicating that the schema was stable for our dataset.

For each sampled evolution commit, we first inspect the pre-change and post-change skill contents together with file-level diffs to identify the modified components.
After the modified components are determined, annotators use each changed component as the annotation unit and assign one \textit{how} label and one \textit{why} label based on the commit message, localized diff, and surrounding skill context.

Based on these manually inspected samples, we first annotate the modified components in each skill evolution event following the same procedure described in Section~\ref{sec:component_label_method}.
After obtaining the component labels, we map the changed lines in each commit to their corresponding components.
Then, for each component, we categorize the corresponding change operations and infer the maintenance intents based on the commit messages, changed lines, and skill content.
We report the annotation reliability for both modified component identification and evolution pattern/motivation labeling in Section~\ref{dis:accuracy}.
The accuracy scores for component, change-pattern, and motivation labels are 0.913, 0.816, and 0.840, indicating that the automated annotations are highly consistent with human labels.
During taxonomy construction, we drew on the classical software maintenance taxonomy~\cite{swanson1976dimensions,iso14764} and adapted it to the agent skill. 

\noindent\textbf{Taxonomy Structure.} 
Figure~\ref{fig:3_taxonomy_evolution} presents the hierarchical structure of our taxonomy for skill evolution.
The taxonomy is organized into four top-level maintenance intents: Corrective, Adaptive, Perfective, and Preventive Maintenance.
Each motivation is further decomposed into fine-grained \textit{why} categories, shown as colored nodes, which are associated with concrete \textit{how} operations, shown as white nodes. 
In this way, the taxonomy connects the reason for a skill change with the specific operation through which the change is implemented.

\textit{Corrective Maintenance} refers to changes that resolve deviations between a skill's current behavior and its intended design goal.
Such changes address missing, incorrect, or unstable skill content, including incomplete functionality, erroneous workflow steps, inaccurate knowledge descriptions, and underspecified execution conditions.

\textit{Adaptive Maintenance} captures changes motivated by external shifts rather than defects in the skill itself.
These changes are needed when the surrounding toolchain, execution environment, interface, data format, or target task evolves, requiring the
skill to be realigned with its updated usage context.

\textit{Perfective Maintenance} describes changes that improve a skill that is already functional.
These changes go beyond basic correctness by enhancing skill capability, efficiency, and usability, reflecting developers' efforts to make skills
more effective in supporting future task execution.

\textit{Preventive Maintenance} denotes changes that proactively reduce future risks or maintenance costs.
Although the current skill may remain executable, it may expose potential security, robustness, traceability, artifact-management, or evaluation
risks. 
Preventive changes aim to mitigate such risks before they lead to failures.

A detailed explanation of the fine-grained categories is provided in our online appendix\footref{fn:online_appendix}.

\noindent\textbf{Automatic Annotation.} Based on this taxonomy, we implement an automatic annotation pipeline. 
Consistent with Section~\ref{sec:3_component_identification}, we also use \texttt{DeepSeek-V4-Flash} as the annotation model. 
To avoid the coarse granularity of commit-level annotation and the excessive fragmentation of line-level annotation, we use the changed component defined in \ref{sec:3_component_identification} as the basic annotation unit. 
Specifically, each changed component is represented as the tuple $\langle \textit{commit}, \textit{file scope}, \textit{component} \rangle$.
For changed components in \texttt{SKILL.md}, we provide the model with commit metadata, the target component, and block-level diffs with the full file as context. 
The block-level diffs include component type, heading path, and localized diffs. 
For changed components in auxiliary files, we extract the corresponding per-file diff from the original commit patch and provide it together with the file path and component label.

Based on the taxonomy and the above artifacts, the model is required to produce a strictly structured JSON output containing one \textit{how} label, one \textit{why} label, and the rationale for the selected labels.
In detail, the model needs to select labels from the predefined taxonomy and return exactly one \textit{how} and one \textit{why} category. 
When the predefined taxonomy is insufficient to explain the change, the model is allowed to suggest a new category as an extension. 
The detailed prompt for annotation is provided in our online appendix\footref{fn:online_appendix}.

\section{Results}

\subsection{RQ1: The Current Status of Skill Evolution}

\subsubsection{RQ1.1: Main Components Distribution}

\begin{figure}[t]
    \centering
    \includegraphics[width=0.8\linewidth]{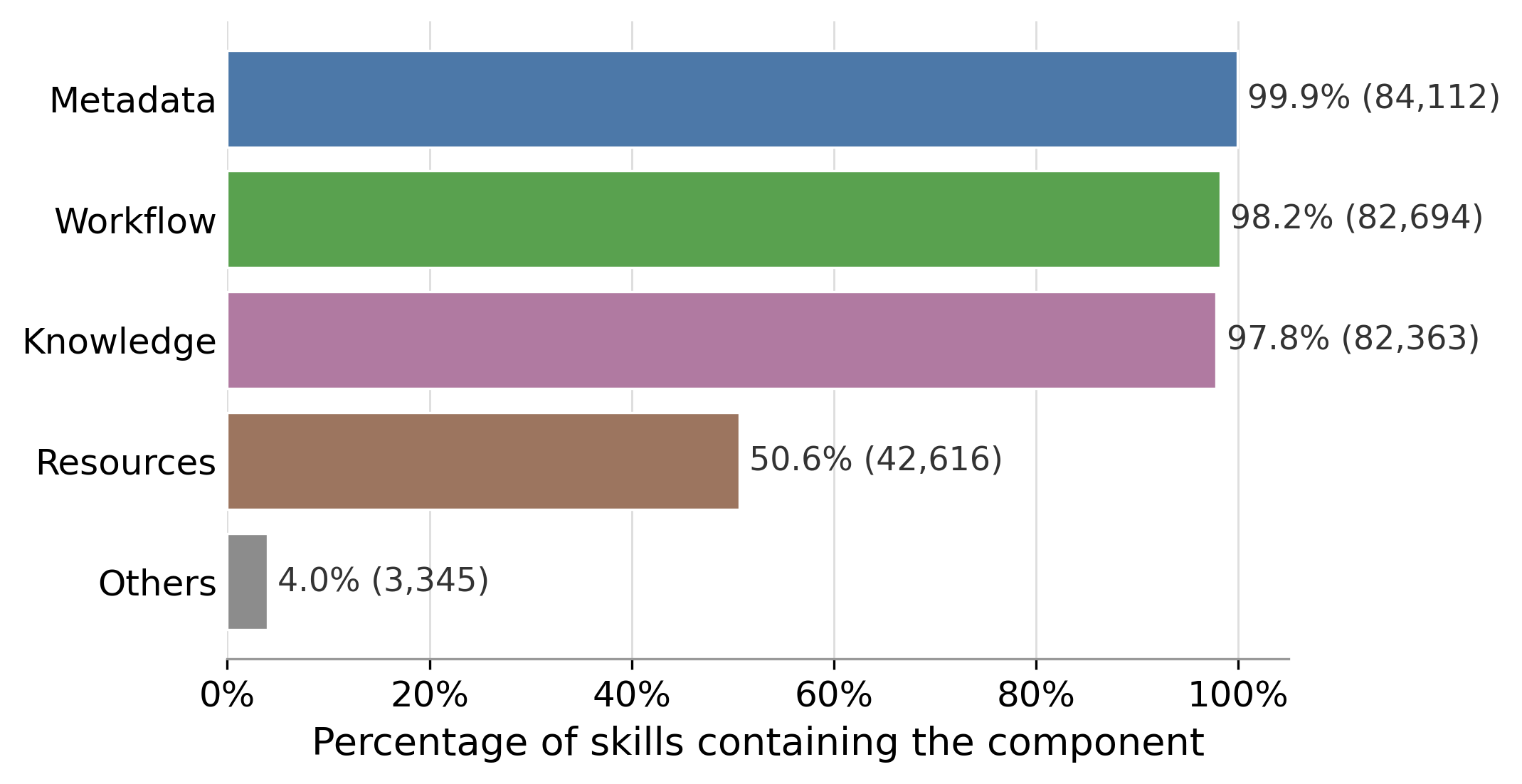}
    \caption{Distribution of Major Skill Components}
    \label{fig:rq11_major_component_distribution}
\end{figure}

\begin{figure}[t]
    \centering
    \includegraphics[width=\linewidth]{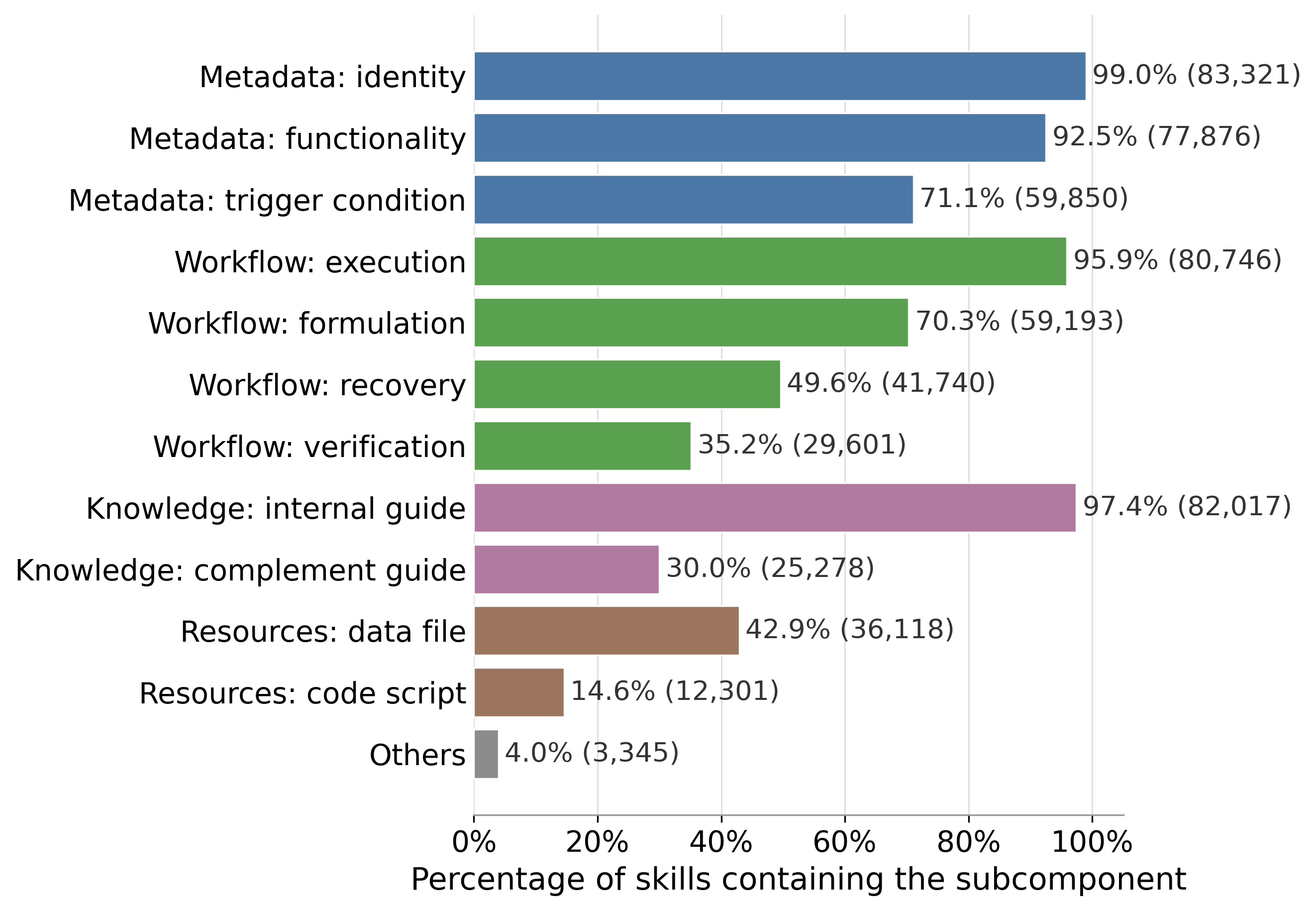}
    \caption{Distribution of Fine-Grained Skill Components}
    \label{fig:rq11_subcomponent_distribution}
\end{figure}

The first research question aims to identify the main functional components that constitute agent skills. 
To answer this question, we apply the component taxonomy defined in Table~\ref{tab:3_skill_component_taxonomy} to the collected skills and quantify the occurrence of each component.  

At the top level, Figure~\ref{fig:rq11_major_component_distribution} shows that the three core components, i.e., metadata, workflow, and knowledge, appear in 99.9\%, 98.2\%, and 97.8\% of skills, respectively. 
In contrast, resources are present in only 50.6\% of skills, with code scripts appearing in only 14.6\% of skills. 
These results suggest that skills are currently used primarily to structure prompts, procedural guidance, and domain knowledge. 
The relatively lower prevalence of code scripts further indicates that many skills do not package custom executable logic, and may instead rely more heavily on natural-language instructions, the agent's general reasoning capabilities, and tools available in the execution environment.

At the fine-grained level, Figure~\ref{fig:rq11_subcomponent_distribution} shows that 92.5\% of skills contain functionality descriptions, while only 35.2\% of skills include verification steps, and only 49.6\% include recovery instructions for handling failures.
This indicates that existing skills often lack support for reliable execution.
A possible explanation is that developers primarily focus on the task specification and execution guidance, implicitly relying on the agent's own execution capability and human intervention to handle failures, leaving verification and recovery overlooked in skill creation.
Moreover, 28.9\% of skills lack trigger conditions and 7.5\% lack functionality descriptions, which may make it harder for agents to retrieve and select.

\answerRQ{1}{
Skills are currently used primarily to structure prompts, procedural guidance, and domain knowledge.
At the fine-grained level, they provide limited explicit quality assurance and failure-handling mechanisms.
36.4\% of skills are incomplete in the metadata component, potentially hindering accurate retrieval by agents.
}

\subsubsection{RQ1.2: Evolution Overview}

\begin{figure}[t]
    \centering
    \includegraphics[width=\linewidth]{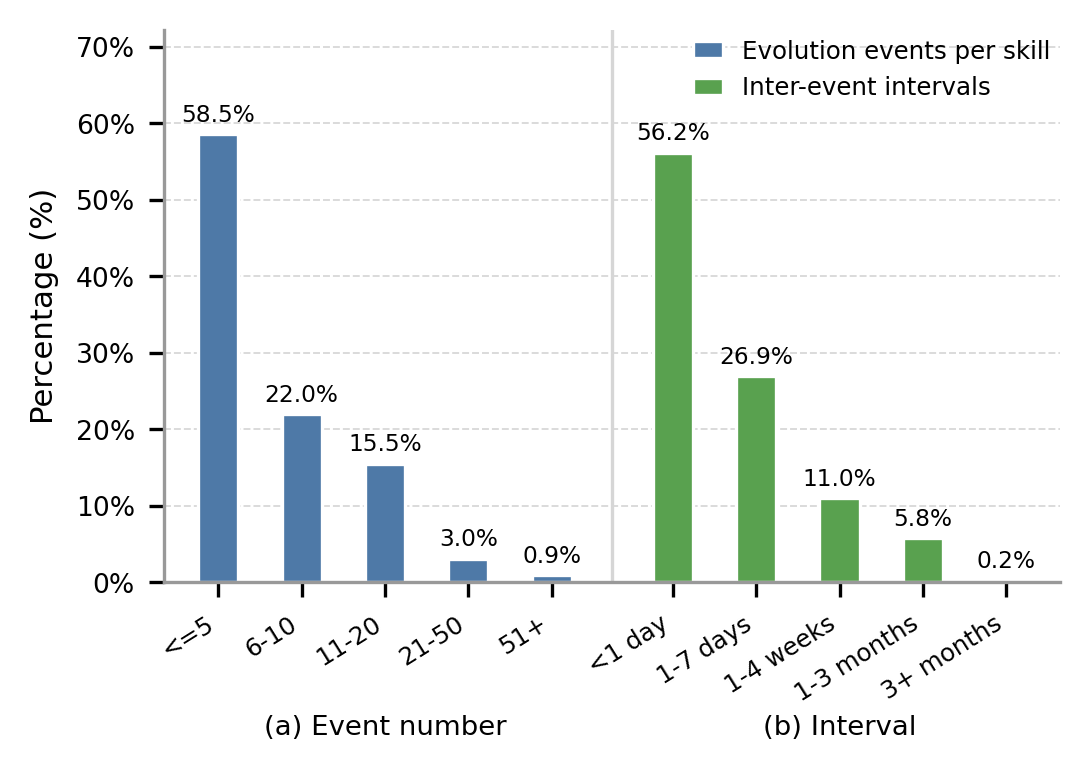}
    \caption{Frequency and Temporal Rhythm of Skill Evolution}
    \label{fig:rq12_evolution_timeline}
\end{figure}

We reconstruct the evolution history of each skill from repository commit histories.
Among the collected skills, 64.3\% have at least two valid commits.
After applying the filtering criteria described in Section~\ref{sec:3_dataset_construction}, there are 84,200 skills with analyzable evolution histories for the following frequency and temporal analyses.

As shown in Figure~\ref{fig:rq12_evolution_timeline}(a), 58.5\% of skills undergo fewer than five evolution events and  4\% of skills undergo more than 20 evolution events, which suggests that most skills experience limited post-creation maintenance.
We further inspect these frequently evolved skills and find that 86.9\% of them are packages that include auxiliary documents, scripts, or project-specific resources.
This pattern reveals that complex and project-coupled skills are more prone to be revised.
As projects frequently update, skills also need to be updated, which imposes a high cost for manual skill maintenance.
Figure~\ref{fig:rq12_evolution_timeline}(b) shows that 83.1\% of consecutive evolution events occur within one week, while only 6.0\% exceed one month.
This indicates that developers often make rapid adjustments based on a skill's early usage feedback until the skill becomes relatively stable.

\answerRQ{2}{
58.5\% of skills experience limited maintenance, and frequent evolution is concentrated in a small subset of complex and project-coupled skills, calling for future developers to focus on them and explore automated maintenance techniques that reduce developers' maintenance cost.
Skill evolution is also temporally concentrated, with most changes occurring within short-term windows.
}

\subsection{RQ2: Modified Components in Skill Evolution}

\begin{figure}[t]
    \centering

    \begin{subfigure}[t]{\linewidth}
        \centering
        \includegraphics[width=\linewidth]{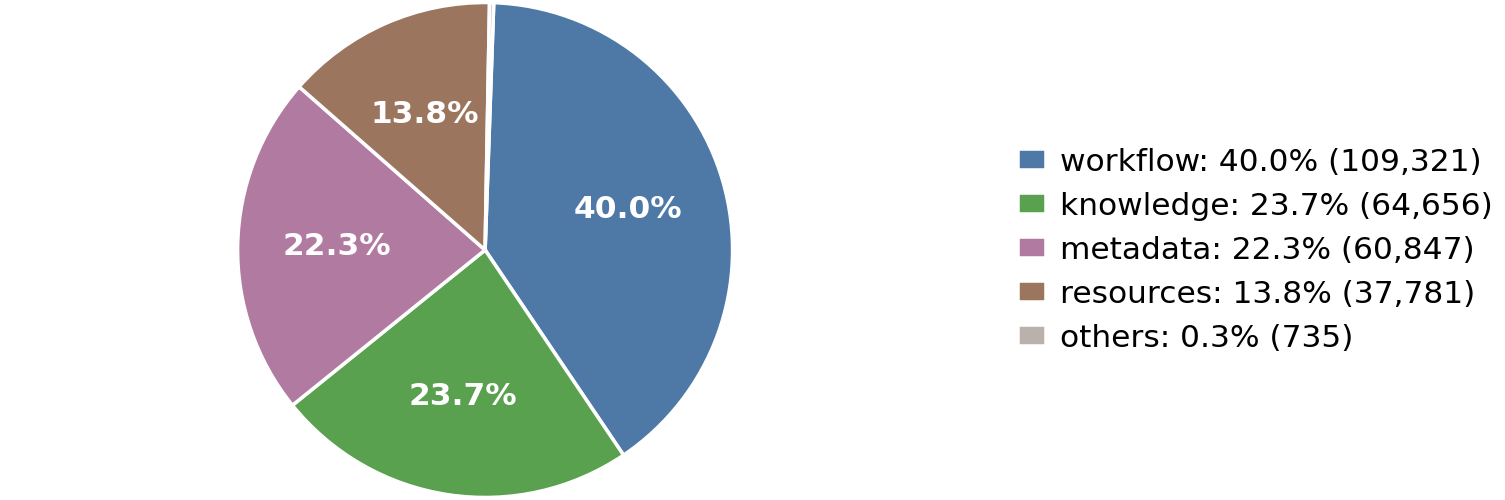}
        \caption{Major Component Distribution}
        \label{fig:rq2_changed_major_components}
    \end{subfigure}

    \vspace{2mm}

    \begin{subfigure}[t]{\linewidth}
        \centering
        \includegraphics[width=\linewidth]{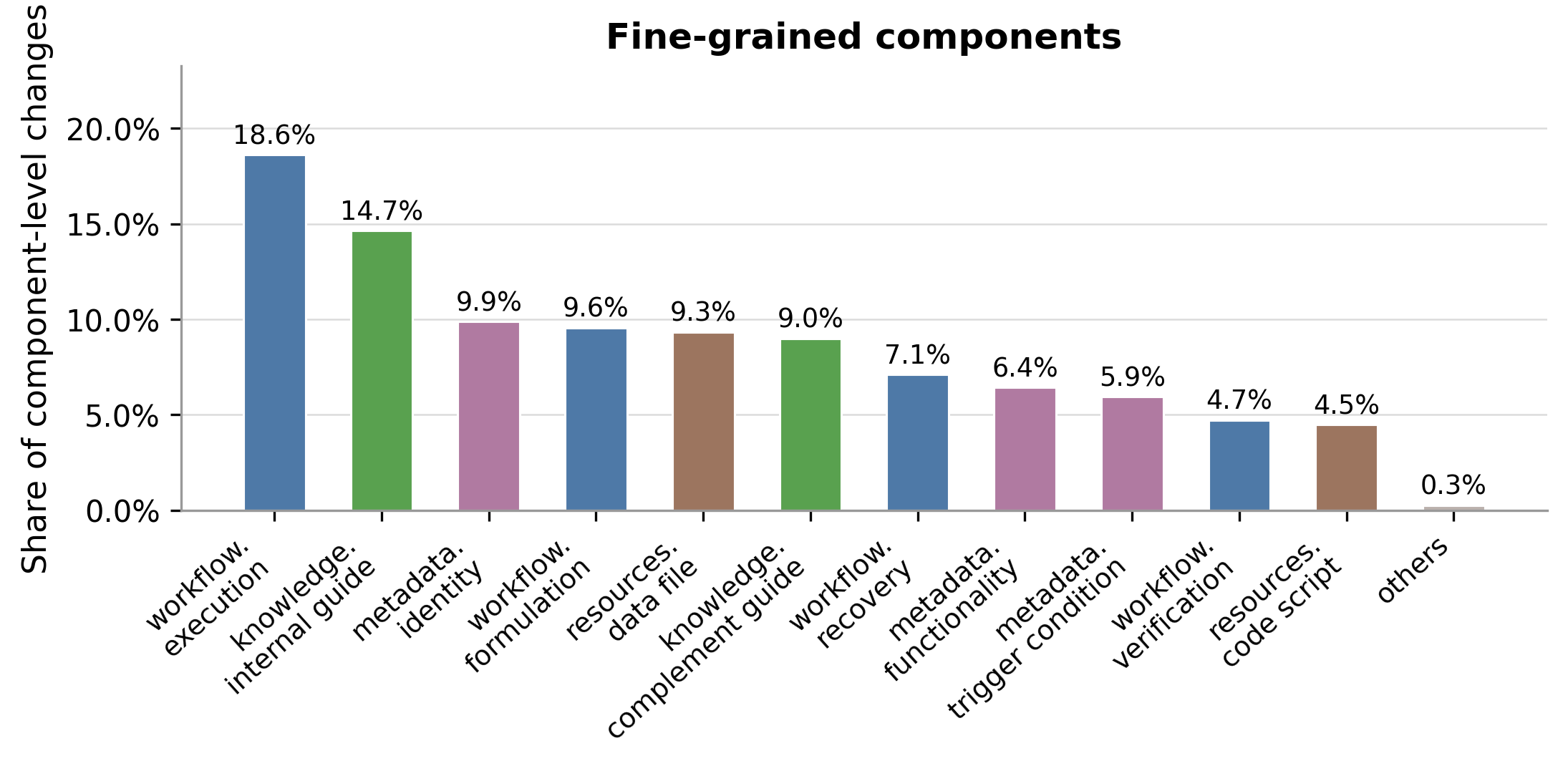}
        \caption{Fine-grained Component Distribution}
        \label{fig:rq2_changed_finegrained_components}
    \end{subfigure}

    \caption{Distribution of Changed Skill Components}
    \label{fig:rq2_changed_components}
\end{figure}

To analyze where skill modifications occur, we apply component-level change analysis described in ~\ref{sec:3_component_identification} to actively evolved skills.
Because applying this analysis to all skills is computationally expensive, we retain skills from repositories with more than 40 stars and skill files with more than 10 commits.
The star threshold filters low-visibility repositories, while the commit-count threshold ensures sufficient evolution information for stable component-level analysis.
This step results in 5,628 skills for the following analysis.

Figure~\ref{fig:rq2_changed_components} presents the distribution of changed major skill components. 
Workflow, knowledge, and metadata together account for 86.0\% of all component-level changes.
Among these components, workflow changes are the most frequent, accounting for 40.0\% of all component-level changes, while resources and others together account for only 14.1\%.
One possible explanation is that resources and others are much less prevalent in skills, appearing in only about half as many skills as workflow (see Figure~\ref{fig:rq11_major_component_distribution}).
Another reason may be that effective workflows usually require iterative refinement. 
This is further supported by the RQ4 result that Workflow Improvement accounts for 18.05\% of all labeled changes.

\answerRQ{3}{
Among actively evolved skills, skill evolution is concentrated in instruction-oriented components.
}

At the fine-grained level, workflow execution (18.6\%) and knowledge internal guidance (14.7\%) are the most frequently modified subcomponents.
The reasons may be that they directly shape agent execution.
In contrast, verification (4.7\%) and recovery (7.1\%) are modified less frequently, suggesting that developers pay comparatively less attention to result-checking and failure-handling logic. 
Given the inherent ambiguity of skills as natural language artifacts and the stochastic nature of LLM-based execution~\cite{ouyang2025empirical, vijayvargiya2025interactive}, the outcomes can be highly unpredictable. 
The reliability of skills is therefore a non-negligible concern for developers.

\answerRQ{4}{
Fine-grained skill evolution mainly targets execution-oriented content, while reliability-oriented components often receive less attention from developers.
}

\begin{table}[t]
  \centering
  \caption{Distribution of Change Patterns.}
  \label{tab:rq3_how_distribution_compact}
  \small
  \begin{tabular}{lr}
  \toprule
  \textbf{How category} & \textbf{Share} \\
  \midrule
  Knowledge Enrichment & 9.00\% \\
  Workflow Refinement & 8.50\% \\
  Traceable Metadata Change & 7.59\% \\
  Knowledge Modification & 7.34\% \\
  Workflow Extension & 6.14\% \\
  Knowledge Alignment & 5.37\% \\
  Format Alignment & 5.01\% \\
  Description Refinement & 4.93\% \\
  Workflow Modification & 4.47\% \\
  Metadata Refinement & 4.46\% \\
  Interface Alignment & 4.15\% \\
  Unnecessary Artifact Cleanup & 3.99\% \\
  Expression Simplification & 3.63\% \\
  Duplication Reduction & 3.44\% \\
  Resource Modification & 2.09\% \\
  Workflow Simplification & 2.07\% \\
  Workflow Alignment & 1.97\% \\
  Content Formalization & 1.89\% \\
  Knowledge Restructuring & 1.74\% \\
  Resource Alignment & 1.44\% \\
  Others & 10.79\% \\
  \bottomrule
  \end{tabular}
\end{table}

\subsection{RQ3: Changed Patterns in Skill Evolution}

Table~\ref{tab:rq3_how_distribution_compact} shows how developers modify skills during evolution.
The most common change operation is Knowledge Enrichment (9.00\%), followed by Workflow Refinement (8.50\%), Traceable Metadata Change (7.59\%), Knowledge Modification (7.34\%), and Workflow Extension (6.14\%), together accounting for 38.57\% of all labeled changes.
In contrast, developers modify skills through structural changes much less frequently.
Knowledge Restructuring (1.74\%), Workflow Integration (0.07\%), and Artifact Refactoring (0.63\%) together account for only 2.44\% of all labeled changes.
The relatively lower prevalence of structural changes suggests that developers are more likely to improve skills through incremental edits than by reorganizing their overall structure.
One possible reason is that such changes are more likely to break or compromise skill functionality and often require more careful consideration.

\answerRQ{5}{
Change patterns are dominated by incremental enrichment and refinement, while structural reorganization is rare.
}
w
It is also noted that simplification is a recurring maintenance behavior in skill evolution.
Unnecessary Artifact Cleanup (3.99\%), Expression Simplification (3.63\%), Duplication Reduction (3.44\%), and Workflow Simplification (2.07\%) together account for 13.13\% of all labeled changes.
For example, \texttt{312362115/skills/deep-research} removed the automatic update-check workflow that repeatedly ran after task completion to reduce redundant execution steps and unnecessary tool overhead.
These changes suggest that developers sometimes encounter context-length pressure issues and attempt to simplify skill content by removing unnecessary artifacts, reducing duplicated instructions, and streamlining workflows.

\answerRQ{6}{
Redundancy is a practical maintenance issue in skill evolution.
Developers can simplify skills by removing unnecessary artifacts, reducing duplicated instructions, and streamlining workflows.
}

\subsection{RQ4: Motivations of Skill Evolution}

\begin{figure}[t]
    \centering
    \includegraphics[width=\linewidth]{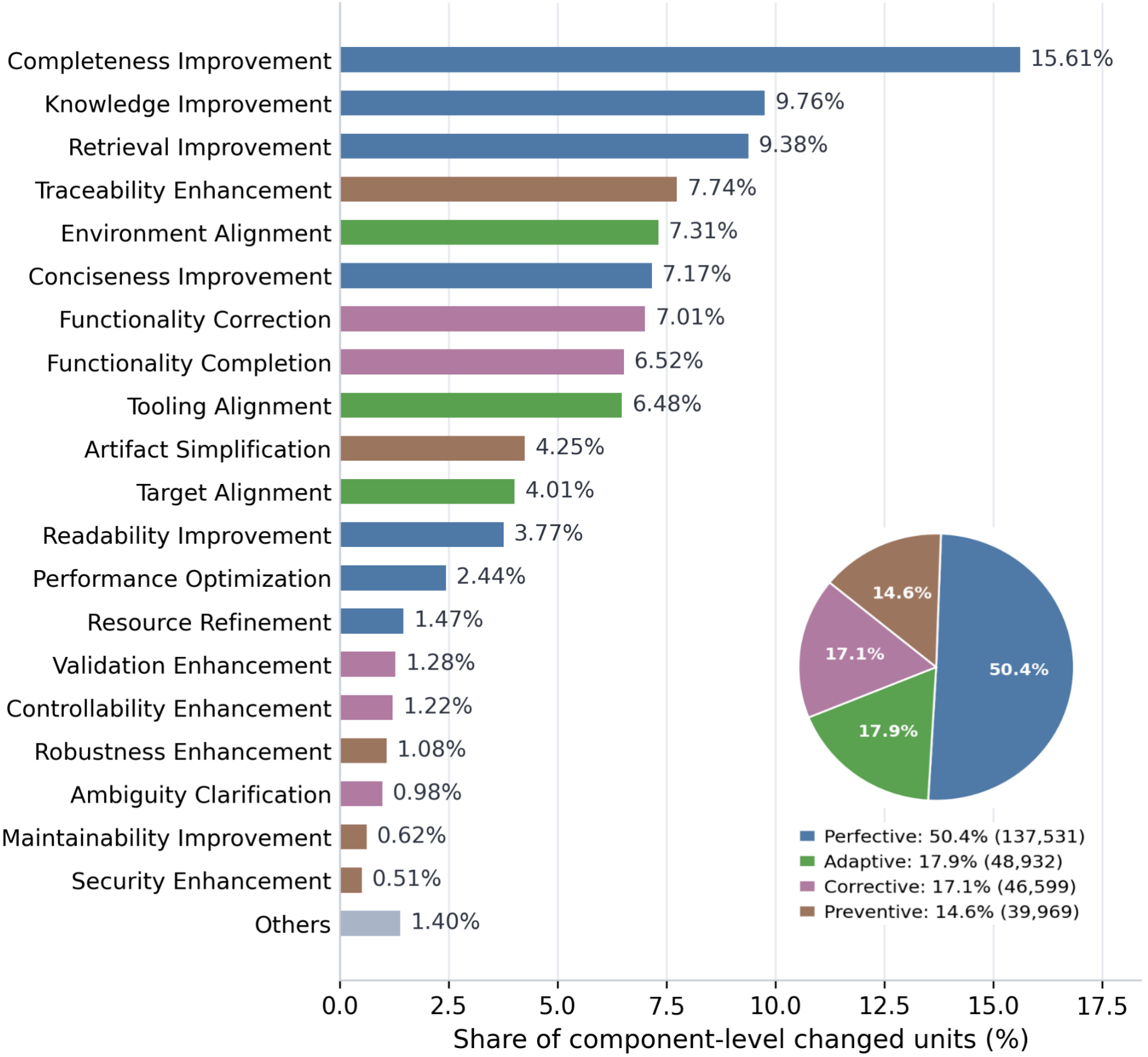}
    \caption{Distribution of Motivations}
    \label{fig:rq4_combined}
\end{figure}

We investigate the intentions behind skill evolution by assigning a motivation label to each modified component (Section~\ref{sec:3_taxonomy_evolution_pattern_and_motivation}), reporting both the top-level distribution and fine-grained categories.

As shown in the pie part of Figure~\ref{fig:rq4_combined}, Perfective maintenance accounts for 50.37\% of all changes, surpassing the sum of the other three types.
This indicates that Perfective maintenance is the dominant motivation for skill evolution. 
The fine-grained motivations further support this tendency. 
Specifically, Completeness Improvement is the most frequent fine-grained motivation, accounting for 15.61\% of all changes, followed by Knowledge Improvement (9.76\%) and Retrieval Improvement (9.38\%), which suggests that skills often require detailed task-related information and clearer retrieval cues after their initial release.
One possible reason is that, unlike traditional software, skills are expressed in natural language with inherently ambiguous correctness boundaries, encouraging developers to continuously refine their skills.
For example, \texttt{0xBigBoss/skills/tilt} added a ``Verifying Updates Applied'' workflow with shell commands to check resource status and wait until resources are ready, thereby improving the completeness of the update process.
\texttt{naksha-studio/skills/design} expanded the trigger condition with visualization keywords, making the skill more likely to be invoked for visualization-related requests.

\answerRQ{7}{
Skill evolution is dominated by Perfective maintenance.
This indicates that skills require continuous refinement to become functional, accurate, and invocable for agents after initial creation.
}

\begin{table}[h]
    \centering
    \caption{Scanner Results for Security Enhancement commits after manual filtering.}
    \label{tab:rq4_security_enhancement_scanner_results}
    \small
    \setlength{\tabcolsep}{4pt}
    \begin{tabular}{lrrrr}
    \toprule
    \textbf{Scanner} & \textbf{Commits} & \textbf{Decreased} & \textbf{Unchanged} & \textbf{Increased} \\
    \midrule
    Cisco & 876 & 133 (15.2\%) & 633 (72.3\%) & 110 (12.5\%) \\
    Snyk  & 876 & 95 (10.8\%)  & 717 (81.8\%) & 64 (7.3\%) \\
    \bottomrule
    \end{tabular}
\end{table}

Moreover, we observe that security-related motivations appear relatively infrequently.
To further contextualize this category, we conduct an auxiliary analysis of changes labeled as \textit{Security Enhancement}, examining whether these security-motivated commits coincide with observable reductions in scanner-reported risks. 
We manually check the 1,156 Security Enhancement commits and retain 876 valid cases for analysis.
Specifically, we use two publicly available skill security scanners, Cisco Skill Scanner~\cite{cisco_skill_scanner} and Snyk Agent Scan~\cite{snyk_agent_scan}, to scan for potential risks in skills.
We analyze their outputs separately and use the number of reported high or critical risks as the main comparison metric.
For each commit, we compare this number before and after the change: a commit is classified as \textit{Decreased}, \textit{Unchanged}, or \textit{Increased} when the number of high-or-critical risks becomes smaller, remains the same, or becomes larger after the commit, respectively.

Table~\ref{tab:rq4_security_enhancement_scanner_results} shows that only a small portion of Security Enhancement commits are accompanied by a reduction in high-or-critical risk findings, with an average reduction rate of 13.01\% across the two scanners.
In contrast, most commits remain unchanged, and a few commits are even associated with an increase in risk findings. 
This suggests that most security-related changes do not lead to security improvements observable by existing skill-security scanners.

\answerRQ{8}{
Security-related skill changes rarely translate into observable scanner-confirmed risk reduction.
This highlights a gap between developers' security intent and actual security outcomes, underscoring the need for more effective automated tools to help developers make substantive improvements to skill security.
}

\section{Discussions}

\subsection{Automated Annotation Accuracy}
\label{dis:accuracy}

As described in Sections~\ref{sec:3_component_identification} and~\ref{sec:3_taxonomy_evolution_pattern_and_motivation}, we first constructed the component taxonomy and
the evolution pattern/motivation taxonomy based on manually inspected samples, and then used LLMs for large-scale automated annotation. To examine the reliability of the automated annotations, we report the manual annotation results for components, change patterns, and motivations, respectively, and cross-validate them against the model-generated annotations. 
Due to space limitations, we only report accuracy scores. Detailed results are provided in our online appendix\footref{fn:online_appendix}.

\begin{table}[t]
    \centering
    \caption{Agreement between automated annotations and manual annotation labels. The first \textit{Component} is sampled skills in RQ1.1, and the latter is the sampled skills in RQ2-4.}
    \label{tab:annotation_accuracy}
    \small
    \setlength{\tabcolsep}{3pt}
    \begin{tabularx}{\linewidth}{>{\raggedright\arraybackslash}X c c c}
    \toprule
    \textbf{Task} & \textbf{Unit} & \textbf{N} & \textbf{Accuracy} \\
    \midrule
    Component (Section~\ref{sec:3_component_identification}) & Skill & 383 & 0.881 \\
    Component (Section~\ref{sec:3_taxonomy_evolution_pattern_and_motivation})
    & Skill & 27 & 0.913 \\
    Change pattern (\textit{how}) & Commit & 404 & 0.816 \\
    Fine-grained motivation (\textit{why}) & Commit & 404 & 0.840 \\
    \bottomrule
    \end{tabularx}
\end{table}

For component annotation, accuracy is computed at the Markdown-block level: we align blocks between the manual annotation and the model annotation within the same skill, and compare their assigned component labels. 
For change pattern and motivation annotation, we compute accuracy over changed components within each commit. 
Specifically, we first align changed components between the reference and model annotations.
Then, for each aligned changed component, the ``how'' or ``why'' prediction is counted as correct only when it matches the reference label.
Changes associated with unmatched components are treated as annotation errors.
The reported accuracy is averaged over commits.

As shown in Table~\ref{tab:annotation_accuracy}, the component annotation achieves an accuracy of 0.881, while the \textit{how} and \textit{why} annotations achieve accuracy scores of 0.816 and 0.840, respectively. 
These results indicate high agreement between model annotations and human annotations, supporting the reliability of the large-scale annotation results used in our analysis.

\subsection{Research Opportunities}
Based on our empirical findings, we summarize the following research opportunities:

\noindent\textbf{Execution-contract skill design.}
RQ1 and RQ2 show that verification and recovery are both underrepresented in current skills and their evolution.
This suggests that many skills still describe task procedures without specifying how agents should check results or handle failures.
Due to the inherent ambiguity of skills as natural language artifacts and the stochastic nature of LLM-based execution, the outcome and procedure can be highly unpredictable. 
Future work could move beyond instruction-only skill design and define execution-contract schemas that include expected outputs, validation criteria, and recovery actions.

\noindent\textbf{Automating just-in-time skill updating.}
RQ1.2 shows that frequent skill evolution is concentrated in a small subset of complex and project-coupled skills.
As these skills are closely tied to project-specific contexts, they must be continuously updated to remain consistent with evolving codebases, dependencies, and tool interfaces.
Such continuous synchronization can impose substantial maintenance costs when performed manually.
These findings highlight the need for automated just-in-time skill updating mechanisms that can detect relevant project changes and adapt associated skills accordingly, thereby reducing developers' maintenance burden.

\noindent\textbf{Context-efficient skill representation.}
Simplification-related changes in RQ3 show that redundancy and accumulated complexity are practical maintenance issues.
At the same time, every loaded skill competes with task history, observations, tool outputs, and retrieved documents for a limited agent context.
This motivates compression and refactoring techniques that remove redundant or irrelevant content while preserving the semantics required for reliable execution.

\noindent\textbf{Retrieval-aware skill engineering.}
Skill retrievability emerges as a recurring maintenance concern.
RQ4 identifies Retrieval Improvement as one of the most frequent motivations.
However, RQ1.1 shows that 28.9\% of skills lack trigger conditions and 7.5\% lack functionality descriptions, which may limit the precise retrieval of skills for a given task.
Future work could build on our dataset to study automated techniques for improving skill retrievability, such as trigger-condition generation, functionality-description refinement, retrieval-keyword suggestion, and related-skill linking.

\noindent\textbf{Context-alignment monitoring.}
RQ4 shows that a significant portion of changes aims to align skills with evolving execution contexts, involving updates to knowledge, formats, interfaces, workflows, and resources.
This reflects the tight coupling between skills and their surrounding execution contexts, such as tools, data formats, and APIs.
A promising direction is to monitor these contextual dependencies and detect when a skill becomes stale or inconsistent with the execution environment.

\noindent\textbf{Skill regression testing.}
The scanner-based analysis in RQ4 shows a gap between security intent and security effect: security-motivated commits do not always reduce risks, and some even increase them.
This calls for regression testing methods that evaluate skill behavior before and after a change.
Possible tests include replaying representative tasks and checking whether the success rate, safety, retrieval accuracy, or execution cost degrades.

\noindent\textbf{Pattern-guided skill enhancer.}
RQ4 shows that skill evolution is dominated by Perfective maintenance, indicating that many skills are functional but still incomplete, under-specified, or difficult to use effectively.
This creates an opportunity for pattern-guided skill enhancers.
Such tools could diagnose weak components, such as missing trigger conditions, sparse verification steps, incomplete recovery logic, or insufficient domain knowledge, and recommend component-level edits based on common evolution patterns.

\subsection{Skill Ecosystem Evolution Trends}

Agent skills are moving from isolated markdown files toward shared and maintained capability artifacts.
Recent skill markets and community repositories suggest a trajectory similar to software ecosystems: from ad hoc scripts, to reusable components, and eventually to maintained ecosystems.
In such an ecosystem, skill evolution is not simply editing documentation, but the basis for keeping skills usable, reliable, and accountable as tasks, tools, and environments change.
Our study shows that current skill evolution is active but still immature, with many changes focusing on refinement and completeness, while verification, recovery, long-term feedback, and security remain underdeveloped.
This suggests that future skill ecosystems need maintenance infrastructure built around four properties: skills should be \textit{triggerable} for
correct retrieval, \textit{attributable} in terms of why they change, \textit{localizable} when failures occur, and \textit{traceable} across their
evolution history.
By characterizing what, how, and why skills evolve, our work provides an empirical foundation for building such reliable and maintainable skill ecosystems.

\section{Threats to Validity}

\noindent\textbf{Internal Validity.}
Our component-level analysis relies on model-based identification of components and their subsequent ``how''/``why'' annotations.
To mitigate this, we first constructed the component and evolution taxonomies from manually inspected samples, used structured prompts with predefined label spaces, and evaluated automated annotations against manually labeled reference data.
The annotation accuracy scores for component, change-pattern, and motivation labels are 0.913, 0.816, and 0.840, respectively, suggesting that the automated annotations are reasonably reliable.

\noindent\textbf{External Validity.}
Our dataset is collected from public GitHub repositories, and the findings may not fully generalize to private skills or enterprise settings. 
Nevertheless, GitHub is the largest open-source code hosting platform, and public agent skill repositories represent the mainstream practices of the developer community. 
In addition, the findings of RQ2--RQ4 are based on relatively active and visible skills, and may not fully generalize to low-activity skills or skills from less visible repositories. 
Nevertheless, skills with fewer evolution events provide limited evidence for fine-grained evolution analysis; focusing on actively evolved skills ensures sufficient change samples for reliable component-level analysis.

\section{Conclusion}

This study systematically analyzes the evolution characteristics of real-world agent skills.
We present the current status of skills and conduct a fine-grained investigation along three dimensions: modified components, change patterns, and evolution motivations.
Using data mining, open coding, semantic annotation, and statistical analysis, we identify several findings.
Agent skills currently lack sufficient attention to reliability and failure handling in both their composition and evolution. 
Evolution changes concentrate in a small subset of project-coupled skills, while modified content skews toward instruction-oriented and perfective improvements. 
Change patterns are dominated by incremental enrichment and refinement, and redundancy emerges as a recurring maintenance issue. 
Security-motivated changes rarely translate into observable risk reduction, revealing a persistent gap between developer intent and actual security outcomes.
These findings also motivate several future research opportunities: execution-contract skill design, automated skill update and refinement, skill regression testing, context-efficiency skill representation, retrieval-aware engineering, and context monitoring.
By releasing our research dataset, future work could develop techniques for improving the quality, reliability, and maintainability of agent skills.

\section{DATA AVAILABILITY}

Our data, code, and online appendix are available at \url{https://github.com/ZJU-CTAG/SkillEMP}.

\section*{Acknowledgment}

This research is supported by the National Natural Science Foundation of China (No.92582107), the Fundamental Research Funds for the Central Universities (No. 226-2025-00067), and the University Grants Committee / Research Grants Council (General Research Fund 16210725 and Theme-based Research Scheme T41-517\_25-N).

\bibliographystyle{IEEEtran}
\bibliography{refs}

\end{document}